\documentclass[sigconf]{acmart}
\AtBeginDocument{%
  }
\setcopyright{acmlicensed}
\copyrightyear{2018}
\acmYear{2018}
\acmDOI{XXXXXXX.XXXXXXX}
\acmConference[EASE 2025]{The 29th International Conference on Evaluation and Assessment in Software Engineering}{17–20 June, 2025}{Istanbul, Türkiye}

\acmISBN{978-1-4503-XXXX-X/2018/06}

\usepackage{package}

\begin{document}

\title{A Catalog of Fairness-Aware Practices in Machine Learning Engineering}

\author{Gianmario Voria}
\email{gvoria@unisa.it}
\affiliation{%
\institution{University of Salerno}
\country{Italy}
}
\author{Giulia Sellitto}
\email{gisellitto@unisa.it}
\affiliation{%
\institution{University of Salerno}
\country{Italy}
}
\author{Carmine Ferrara}
\email{carferrara@unisa.it}
\affiliation{%
\institution{University of Salerno}
\country{Italy}
}
\author{Francesco Abate}
\email{f.abate20@studenti.unisa.it}
\affiliation{%
\institution{University of Salerno}
\country{Italy}
}
\author{Andrea De Lucia}
\email{adelucia@unisa.it}
\affiliation{%
\institution{University of Salerno}
\country{Italy}
}
\author{Filomena Ferrucci}
\email{fferrucci@unisa.it}
\affiliation{%
\institution{University of Salerno}
\country{Italy}
}
\author{Gemma Catolino}
\email{gcatolino@unisa.it}
\affiliation{%
\institution{University of Salerno}
\country{Italy}
}
\author{Fabio Palomba}
\email{fpalomba@unisa.it}
\affiliation{%
\institution{University of Salerno}
\country{Italy}
}

\renewcommand{\shortauthors}{Voria et al.}

\keywords{Fairness, AI Development Lifecycle, Software Engineering for AI.}

\begin{abstract}
Artificial Intelligence (AI)'s widespread adoption in decision-making processes, particularly with the introduction of AI-based assistants, raises concerns about ethics and fairness, particularly regarding the treatment of sensitive features and potential discrimination against underrepresented groups. The software engineering community has responded by developing fairness-oriented metrics, empirical studies, and mitigation approaches. However, there remains a gap in understanding and categorizing practices for engineering fairness throughout the development lifecycle of AI-based solutions.  
This paper presents a catalog of practices for addressing fairness derived from a systematic mapping study. The study identifies and categorizes 28 practices from existing literature, mapping them onto different stages of the development lifecycle. From this catalog, actionable items and implications for both researchers and practitioners in software engineering were extracted.
This work aims to provide a comprehensive resource for integrating fairness considerations into the development and deployment of AI systems, enhancing their reliability, accountability, and credibility. 
\end{abstract}

\maketitle
\section{Introduction}
Artificial Intelligence (AI) and Machine Learning (ML) have become integral to decision-making processes and task automation across various domains \cite{zhou2018human}. AI-enabled systems, software systems that include at least one component powered by AI algorithms \cite{ai_enabled}, have demonstrated their effectiveness in numerous critical areas. However, ensuring the \textit{quality} of these systems extends beyond their accuracy and efficiency—\textbf{fairness} is a fundamental property, particularly for AI systems that generate recommendations or influence human decisions \cite{mehrabi2021survey}. The risk of bias in machine learning models remains a pressing challenge, as they can produce discriminatory outcomes against minorities due to biases present in historical training data. Several high-profile incidents have underscored the importance of fairness in AI systems \cite{ia_ethical_incidents}. For example, a Facebook vision model inappropriately labeled images of Black men as "primates," Amazon's algorithm unfairly lowered the sales rankings of books with LGBTQ+ themes, and authentication processes have exhibited biases based on race and gender. \revised{Moreover, with the recent introduction of AI-based assistants based on large language models (LLMs) \cite{llm_assisted_se, copilot_eval}, such as Github's CoPilot and ChatGPT, concerns regarding their ethical implications have emerged. Kotek et al. \cite{Kotek_2023} tested AI-assistants to evaluate the fairness of their responses, revealing gender-related biased assumptions influenced by subjective opinions rather than factual evidence. Also, Truede et al. \cite{treude2023elicits} found notable gender bias in software development tasks like issue assignment, highlighting the need for improved training in LLMs}. These cases highlight how fairness is not just a desirable feature but an essential quality attribute for AI-enabled systems, particularly those interacting with or making recommendations to users.

Recognizing these ethical concerns, the research community increasingly treats ethics and fairness as critical non-functional properties of ML-enabled systems \cite{brun2018software}. \revised{One approach to enhancing fairness has focused on \textit{bias mitigation algorithms}, i.e., algorithmic solutions designed to address bias in AI models \cite{10.1145/3631326}. While these methods have demonstrated improvements in fairness, they have also been found to impact other non-functional requirements, such as accuracy or energy efficiency \cite{chen2024fairness, de4966447examining}. Furthermore, recent research has highlighted that their adoption remains limited due to practical challenges in their application \cite{toolkit_landscape, toolkit_survey}.} To complement these efforts, researchers have also proposed various engineering practices to integrate fairness throughout the development pipeline, including fairness-aware training \cite{li2022training}, hyper-parameter optimization \cite{chakraborty2019software}, and automated fairness testing \cite{galhotra2017fairness}. \revised{Instances of these practices have been explored across different types of AI-enabled systems, from balancing sensitive features in datasets for training ML models \cite{chakraborty2021bias} to fairness testing techniques for natural language-based systems \cite{asyrofi2021biasfinder}. We refer to these approaches as \textit{fairness-aware practices}, i.e., development strategies that can be adapted to enhance fairness.} Despite the \textit{promising impact of fairness-aware practices}, our analysis reveals that most research has centered on algorithm selection and bias mitigation techniques at the training stage, often overlooking fairness considerations at the design phase or in the context of evolving systems \cite{ferrara2024fairness}. 

In this paper, we introduce a catalog of 28 fairness-aware practices that practitioners can adopt to address fairness as an integral aspect of AI system development. Our catalog aims to bridge the gap between isolated interventions and holistic strategies, providing a structured framework that supports fairness throughout the entire AI system development life cycle. Finally, we provide an actionable \textit{research roadmap} elicited from the findings of our study, with a particular focus on how to design and implement \textit{fairer} AI-based solutions or assistants and how to design recommendation systems based on this catalog to help practitioners in developing fairer solutions.

\vspace{5pt}

\stesummarybox{\faBullseye \hspace{0.05cm} Research Goal.}{The objective of our research is to elicit from literature a catalog of fairness-aware practices to (1) guide practitioners in the \textbf{development of fairer AI}-enabled solutions throughout the whole life cycle and (2) enable further research to \textbf{design new recommendation systems} to support practitioners in the selection of fairness-aware practices.}

\vspace{5pt}

To develop the catalog, we conducted a systematic mapping study to synthesize existing knowledge on fairness-aware practices proposed by the research community. In this paper, we present these practices, provide concrete examples that practitioners can apply to address fairness concerns in real-world scenarios, and outline future directions for AI fairness in software design, maintenance, and evolution. Through this work, we offer actionable guidance for integrating fairness into Software Engineering for Artificial Intelligence.
\section{Background}
With the popularity of Artificial Intelligence constantly increasing, the Software Engineering research community has been proposing novel techniques and methods to make the design, development, and verification of ML-enabled solutions more effective~\cite{belani2019requirements}. This is what has been coined as \emph{Software Engineering for Artificial Intelligence} (SE4AI).
One of the most relevant contributions introduced by the SE4AI research community concerned the introduction of specific strategies and models to engineer the development and maintenance processes of ML-enabled software systems. Burkov~\cite{burkov2020machine} pioneered the work in this respect by proposing a life cycle model for machine learning systems, which includes a number of steps, such as (1) Objectives definition, (2) Data collection and preparation, (3) Feature engineering, (4) Model training, (5) Model Deployment, (6) Operating phase and monitoring, and (7) Maintenance and evolution of the model.

\revised{However, there is a lack of appropriate instruments to address non-functional requirements in AI-based systems, which are essential for ensuring their overall quality. This gap can hinder developers in maintaining and evolving these systems while also causing them to deviate from expected behavior, ultimately affecting user trust and reliability. In this paper, we focus on one critical quality attribute—\emph{fairness}—and explore how it can be effectively integrated into the development and evolution of AI systems.}

\smallskip \textbf{Fairness}
Recent studies have shown that suboptimal design choices in AI systems can lead to discrimination against users~\cite{brun2018software}. Several real-world cases highlight this issue, such as (1) racial bias in medical cost predictions~\cite{obermeyer2019dissecting}, (2) unfair assessments of criminal recidivism~\cite{biasblack2016propublica}, and (3) gender discrimination in automated hiring~\cite{amazonrecruiting2018reuters}. These cases demonstrate that fairness is not just a theoretical concern but a pressing issue in AI deployment.

\emph{Software fairness} encompasses principles, techniques, and practices that ensure AI systems operate ethically. However, over 20 definitions exist, broadly categorized into three main approaches: (1) ensuring equal positive outcomes across groups, (2) treating similar inputs similarly, and (3) preventing unintended causal relationships between features and predictions~\cite{verma2018fairness}.

A key challenge in AI fairness is biased training data, as datasets often contain imbalances that reinforce discrimination~\cite{vasudevan2020lift}. Prior research has emphasized data diversity as a fairness strategy~\cite{moumoulidou2020diverse}, yet studies show that merely increasing dataset features does not necessarily reduce bias~\cite{zhang2021ignorance}. Bias often originates in data selection, influencing the fairness of model outcomes~\cite{chakraborty2021bias}.

Beyond data concerns, Software Engineering research emphasizes fairness throughout the AI lifecycle. Brun et al.\cite{brun2018software} highlight the need for fairness-aware design and tools to detect discrimination, while Finkelstein et al.\cite{finkelstein2008fairness} link fairness issues to requirement engineering flaws, proposing optimization methods to balance conflicting fairness definitions. Other work has explored fairness testing, with automated tools assessing bias in AI models~\cite{galhotra2017fairness}.

While prior research has introduced algorithmic solutions or fairness interventions at specific stages, \emph{our study aims to identify and systematize fairness-aware practices that practitioners use across the AI system lifecycle, bridging the gap between isolated efforts and a holistic approach to fairness}.

\section{Research Method}
The ultimate \emph{goal} of our work was to elicit the most common literature practices to handle fairness in the \emph{context} of AI and ML-enabled systems. The \emph{perspective} is of both researchers and practitioners. The former are interested in understanding the state of practice in the field, with the objective of finding future research directions or current literature gaps. The latter are interested in understanding the practices that can be used to address fairness throughout all the stages of the life cycle.

To reach the goal of extracting a catalog of fairness-aware practices, we conducted a systematic mapping study. Based on the guidelines by Peterson \cite{petersen2008systematic}, the study consisted of three main processes: (1) the study selection, which allowed us to obtain a complete set of primary research studies; (2) the data extraction, by which we analyzed the collected studies to obtain information about the practices; and (3) the catalog formalization, where we conducted iterative content analysis sessions \cite{white2006content}. This resulted in a \emph{validated} catalog of fairness-aware practices mapped to ML life cycle stages. This structured overview informs both practitioners and researchers about fairness practices across ML development.

Among the various models that formalize the phases included within the life cycle of an ML-enabled system, the model proposed by Burkov \cite{burkov2020machine}, is the one that most explicitly describes the main phases that practitioners should conduct to apply it in practice. Hence, we sought to elicit practices that might be employed within each of the six main phases of the AI development pipeline: C1 - Requirement Elicitation \& Analysis, C2 - Data Preparation, C3 - Model Building, C4 - Model Training \& Testing, C5 - Model Verification \& Validation, and C6 - Model Maintenance \& Evolution.

\begin{figure}
    \centering
    \includegraphics[width=.6\linewidth]{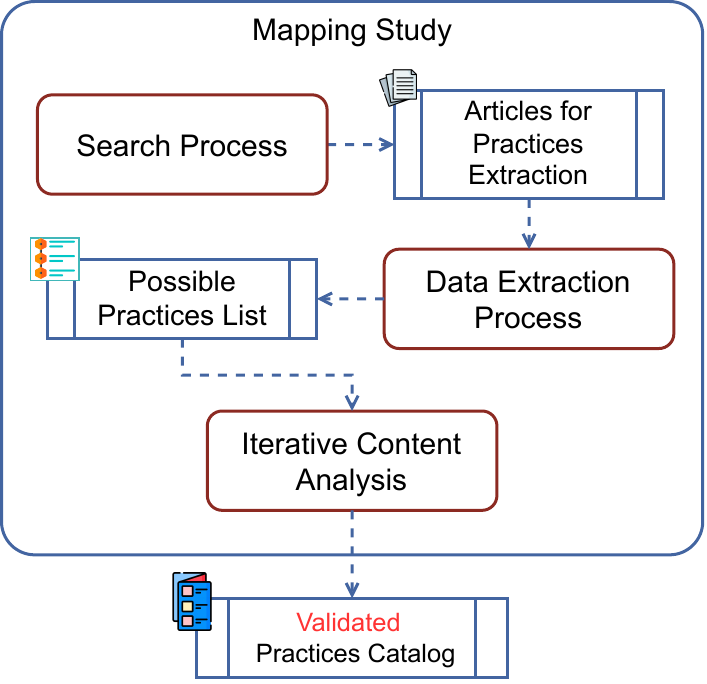}
    \caption{Overview of the methodology to achieve the objective of the study.}
    \label{im_a_methodology}
\end{figure}
\subsection{Search Process Definition}
We selected \textsc{Scopus}\footnote{The \textbf{\textsc{Scopus}} platform: \url{https://www.scopus.com/}.} as the research database for our mapping study. As remarked in the \href{https://www.elsevier.com/solutions/scopus/why-choose-scopus}{platform description}, this is one of the best interdisciplinary research databases and guarantees various analytical tools and APIs to visualize and analyze the search results quickly. As such, it perfectly fitted the scope of our study: fairness is indeed a context-dependent and multidisciplinary topic, and the selection of \textsc{Scopus} could lead us to identify articles coming from various research communities and published in interdisciplinary conferences and journals. 

Following the \href{https://service.elsevier.com/app/answers/detail/a_id/34325/}{recommendations} provided by \textsc{Scopus} on the syntax of the research queries, we decided to formulate a research query that would represent each phase ML life cycle; in this way, we could identify fairness-aware practices for each phase. In particular, we adopted the following strategies to formalize the query: (1) we considered as \textbf{required keywords} \emph{fairness}, \emph{machine learning} and \emph{pipeline} to identify our search theme; (2) we included \textbf{specific keywords} in OR to retrieve papers targeting fairness-aware practices related to the corresponding phase of the pipeline.

A specific engineered development life cycle for AI-enabled systems was proposed in recent years, but the concept of fairness was well-known also before. Hence, we considered it reasonable not to limit our search scope to the last few years.

The next step of our study was the formulation of the exclusion and inclusion criteria. As for the former, we filtered out the resources based on the following criteria:

\begin{itemize}
    \item EC1 - Duplicated resources;
    
    \item EC2 - Papers not written in English;
    
    \item EC3 - Short papers, namely papers that were less than four pages;
    
    \item EC4 - Papers whose access was not allowed;
\end{itemize}

As for the inclusion criteria, we included papers that met the following criteria:

\begin{itemize}
    \item IC1 - Methodological or empirical papers concerning machine learning/artificial intelligence and/or specific phases of a development pipeline;
    
    \item IC2 - Papers focused on AI fairness;
    
    \item IC3 - Papers that explicitly or implicitly proposed one or more practices to deal with fairness in an ML pipeline.
    
\end{itemize}

While the first two inclusion criteria allowed the identification of primary studies that could be potentially relevant to identify practices, IC3 ensured that the resources identified were actually relevant. According to the guidelines by Wohlin~\cite{snowballing}, we applied the exclusion \& inclusion criteria assessment twice, namely before and after the snowballing step described in the following subsection. 

A \textit{snowballing} process was applied to identify additional resources that were not included through the initial search. To be comprehensive, we applied \emph{backward} snowballing, scanning the list of references of the primary studies that successfully passed the inclusion criteria, while \emph{forward} snowballing was applied considering papers that cite the previous ones.

\begin{figure*}[h]
    \centering
    \includegraphics[width=.8\linewidth]{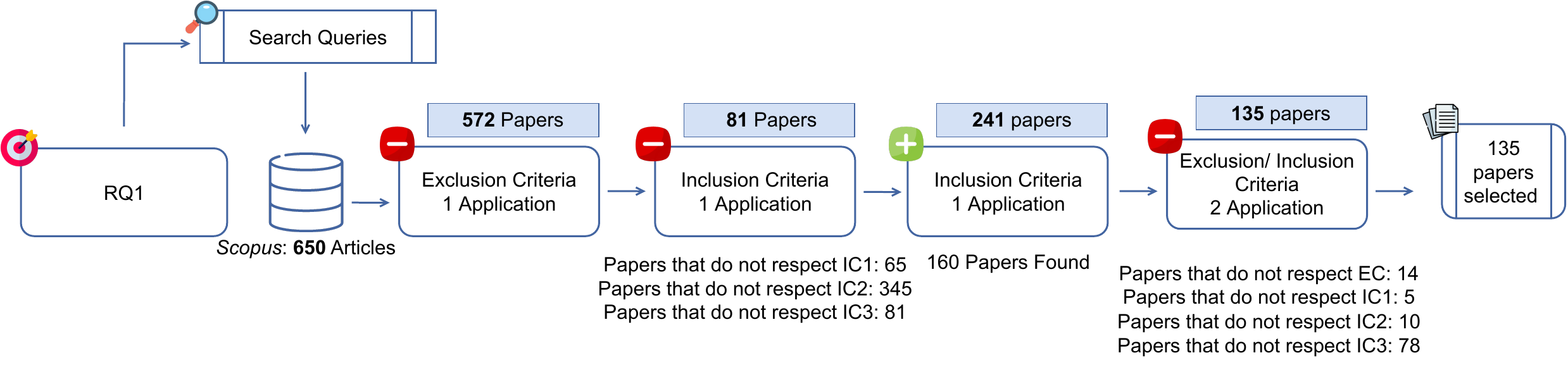}
    \caption{Overview of the Search Process Execution.}
    \label{im_a_mapping}
\end{figure*} 
\subsection{Search Process Execution}
One of the authors (referred to as the \emph{main inspector}) led all phases of the mapping study, with another author (the \emph{assistant inspector}) providing support and double-checking each step. The remaining authors reviewed the process to ensure consistency and reliability, identifying any overlooked concerns. A detailed spreadsheet of the selection process and its outcomes is available in our online appendix \cite{appendix}, while Figure \ref{im_a_mapping} summarizes the search process execution.
Initially, the main inspector's query on the \textsc{Scopus} search engine yielded 650 hits. Information about these papers, including titles, abstracts, and publication details, was compiled into a spreadsheet to facilitate subsequent steps. The main inspector then applied exclusion criteria based on this information. In cases where available information was insufficient, both inspectors examined the full article content to determine its inclusion or exclusion. This process resulted in the removal of 78 articles with the following breakdown:

\begin{itemize}
    \item 4 papers that matched EC1 (duplicated papers);
    \item 4 papers that matched EC2 (non-English papers);
    \item 6 papers that matched EC3 (short papers);
    \item 64 papers that matched EC4 (full-text not available). 
\end{itemize}

The remaining 572 articles underwent further scrutiny. The main inspector applied inclusion criteria based on a full-text reading, which led to the exclusion of 491 papers (65 for not meeting IC1, 345 for IC2, and 81 for IC3). All authors validated the 81 studies that passed this step, approving them for the subsequent snowballing process.

The main inspector then conducted snowballing on these 81 papers, identifying 160 additional resources. To maintain quality, only resources available on \textsc{Scopus} were considered, avoiding unpublished work. Applying exclusion and inclusion criteria to these snowballed resources resulted in the removal of 107 papers (14 due to EC, 5 for IC1, 10 for IC2, and 78 for IC3). This concluded the search process, yielding 135 primary studies for further analysis.

Before extracting and analyzing data to elicit fairness-aware practices, we performed a preliminary classification of the primary studies. Our classification schema included 4 facets: 
\begin{itemize}
    \item The \textsl{'paper type'}, which allowed us to distinguish our resources in: (1) Journal papers, (2) Conference papers, (3) Magazine papers.
    
    \item the \textsl{'research type'}, that allowed us to distinguish between different types of studies, as described by Peterson \cite{petersen2008systematic}: (1) Validation research; (2) Evaluation research; (3) Solution proposal; (4) Philosophical paper; (5) Opinion paper; and (6) Experience paper. The definition of each research type is also reported in our online appendix \cite{appendix}. 

     \item The \textsl{'knowledge area'}, that is, the pipeline stage(s) considered in the study. The values of this facet corresponded to the phases of the incremental-iterative process considered in the study. A study could be assigned to one or more knowledge areas, depending on how many phases were considered.
    
    \item The \textsl{'main topic'} of the paper, which allowed us to distinguish the primary studies in (1) Fairness-related and (2) AI development, including fairness treatment.

\end{itemize}

The classification step was initially conducted by the main inspector and later double-checked by the other authors. Among the various relevant pieces of information collected, it is worth observing that the resources we collected are divided into:
\begin{itemize}
    \item Journal papers (57), respectively divided into 28 solution proposals, 16 experience papers, 5 philosophical papers, 6 evaluation researches, and 2 Opinion papers;
    \item Conference papers (71), respectively divided into 49 solution proposals, 11 experience papers, 2 philosophical papers, 9 evaluation researches;
    \item Review papers (7), respectively divided into 6 experience papers, 1 philosophical paper;
\end{itemize}

Concerning the \textsl{'main topic'} and the \textsl{'knowledge area'}, we classified 81 papers as \emph{mainly focused on fairness} and 54 \emph{generally concerning ML-enabled systems development, involving fairness as an active part of the study. Among them, 16 papers were considered useful to extract fairness treatment information pertaining  \textsl{' Requirements elicitation and analysis'} phase, 55 pertaining \textsl{'Data preparation'}, 44 \textsl{'Model building'}, 38 \textsl{'Model training and testing'}, 51 \textsl{'Model verification and validation'}, and 6 \textsl{'Model maintenance and evolution'}}.

\subsection{Data Extraction and Analysis}
To enhance the effectiveness and replicability of the data extraction process, we first compiled a spreadsheet---available in our online appendix \cite{appendix}---containing all relevant information useful for classifying fairness-related practices. This work involved two steps:

\begin{itemize}
    \item \textsl{'keyword analysis'}: in the full-text of each primary study, one of the inspectors looked for specific keywords such as \emph{''fairness''}, \emph{''bias''}, \emph{''discrimination''}, \emph{''treatment''}, and \emph{''practice''}. This operation allowed for the quicker identification of relevant pieces of information.
    
    \item \textsl{'deep content analysis'}: The involved inspector read the entire content of the paper abstract and the main paragraphs of the methodological part to find useful information that we did not detect with the previous step. 
\end{itemize}

The main and assistant inspectors equally split the data extraction workload by analyzing 68 and 67 papers each. To increase the validity of the extracted data, one inspector double-checked the information extracted by the other and vice versa. Any disagreement was promptly discussed and solved.

The data extracted were then analyzed by the inspectors following a formal iterative content analysis process \cite{white2006content}. More specifically, we conducted three main operations:

\begin{itemize}
    \item \emph{Micro-analysis}: Inspectors manually analyzed extraction forms and labeled practices with similar goals or quality aspects. They focused on the \emph{'Possible practice'} field, with the option to use an \emph{'Not a practice'} label when appropriate. The main inspector conducted the initial analysis, followed by the assistant inspector's review. Other authors then validated and provided feedback on improving or aggregating labels.

    \item \emph{Categorization}: Based on feedback from the first iteration, the main inspector conducted a second iteration to clarify ambiguous labels and cluster practices with identical labels.
    
    \item \emph{Saturation}: The micro-analysis and categorization steps were iteratively performed until reaching theoretical saturation, where all authors accepted the content analysis and no further labels were necessary to represent all concepts correctly.
\end{itemize}

The data extraction process yielded 414 potential practices, averaging 3.06 practices per primary study. The distribution of these practices across the ML pipeline generally aligned with the trends observed in the primary study classification, with some notable variations. The breakdown was as follows: 119 practices for \textsl{'Data Preparation'}, 65 for \textsl{'Model Training and Testing'}, 77 for \textsl{'Model Validation and Verification'}, 109 for \textsl{'Model Building'}, 38 for \textsl{'Requirements Elicitation and Analysis'}, and 6 for \textsl{'Model Maintenance and Evolution'}.
Many initial practices represented similar fairness-handling actions, necessitating refinement through iterative content analysis. This process, involving two rounds of micro-analysis and categorization, resolved 15 instances of disagreement between inspectors through online discussions.

The analysis condensed the initial set to 28 distinct practices, mapped onto the six phases of the incremental-iterative ML pipeline process. The final set of practices and their pipeline mappings are available in a spreadsheet in our online appendix \cite{appendix}.
\section{Results}
\renewcommand{\arraystretch}{1.2}
\begin{table}
\caption{Fairness-Aware Practice Catalog.}
\label{tab:catalog}
\centering
\resizebox{\linewidth}{!}{
\begin{tabular}{|p{8.3cm}|}
\rowcolor{purple}\textcolor{white}{\textbf{C1 - Requirements Elicitation \& Analysis}} \\
 1. Empirical methodologies for fair requirement elicitation and analysis  \\
\rowcolor{purple!20} 2. Multi-objective optimization for fairness constraints \\
 3. Reverse engineering to elicit fairness requirement of a new system  \\

 \rowcolor{purple} \textcolor{white}{\textbf{C2 -  Data Preparation}} \\
\rowcolor{purple!20} 1. Data balancing techniques to respect fairness constraints\\
2. Data mining approaches to discover discrimination\\
\rowcolor{purple!20} 3. Data \& features transformation strategies under fairness constraints\\
4. Diversity data set selection for sensitive groups' representations\\
\rowcolor{purple!20} 5. Causal analysis approaches to identify discrimination dependencies in data\\
6. Measurement approaches to improve data fairness under multiple quality constraints\\
\rowcolor{purple!20} 7. Multitask learning to maximize minority groups' representativeness before training\\

  \rowcolor{purple}\textcolor{white}{\textbf{C3 - Model Building}} \\
1. Ensemble learning strategies under different fairness definitions and constraints \\
\rowcolor{purple!20} 2. Focused learning strategies to obtain discrimination-free outcomes \\
3. Fair regularization terms according to specific fairness metrics and constraints \\
\rowcolor{purple!20} 4. Adversarial learning strategies to balance fairness in quality trade-offs of the model \\

  \rowcolor{purple} \textcolor{white}{\textbf{C4 - Model Training \& Testing}} \\
1. Fairness hyper-parameters tuning\\
\rowcolor{purple!20} 2. Post-processing transformation to balance results among minority groups\\
3. Post-processing strategies to optimize fairness levels of the system\\
\rowcolor{purple!20} 4. Fair test suites generation strategies\\
5. Mutation testing for unfairness causes the detection\\
\rowcolor{purple!20} 6. Testing strategies based on correct prediction oracles\\

  \rowcolor{purple} \textcolor{white}{\textbf{C5 - Model Verification \& Validation}} \\
1. Validation strategies to detect discrimination according to different meanings of data\\
\rowcolor{purple!20} 2. Features causal dependencies analysis to remove causes of discrimination\\
3. Model comparisons for fairness levels improvement\\
\rowcolor{purple!20} 4. Definitions of fairness validation strategies among different definitions and metrics\\
5. Formal validation strategies to evaluate and improve fairness trade-offs\\

  \rowcolor{purple} \textcolor{white}{\textbf{C6 - Model Maintenance \& Evolution}} \\
\rowcolor{purple!20} 1. Feature standardization to improve fairness\\
2. Model outcomes analysis to improve fairness\\
\rowcolor{purple!20} 3. Multiple Datasets analysis to improve sensitive data representativeness\\
\hline

\end{tabular}
}
\end{table}

In this section, we discuss the main findings of our research, i.e., the catalog of practices to deal with fairness. In particular, we depict the practices identified in the mapping study, along with their categorization within the phases of the AI development pipeline.  For each of the six categories, we report the number and kind of research article that discusses the identified practices. Table \ref{tab:catalog} summarizes the whole catalog of fairness-aware practices elicited.

To support practitioners in their adoption of the fairness-aware practices in the catalog, we provide a practical scenario in which the practice could be applied. The scenario is tailored on an unfair system, in particular the well-known \textsl{COMPAS} case study~\cite{biasblack2016propublica}, i.e.,  a system based on machine learning aimed at predicting whether arrested people are likely to reiterate with their crimes over the years, that has been demonstrated to be discriminating.

The complete data analysis spreadsheets are available in the online appendix of this paper~\cite{appendix}. Due to space limitations, the complete set of references for all the practices is also detailed in our online appendix.

\subsection{C1 - Requirements Elicitation \& Analysis}
The first category of the catalog includes the three identified practices that are related to the first step of an AI-enabled system pipeline, i.e., requirements elicitation and analysis. A total of 16 papers mentioned one or more of the practices in this category; nine of them were specifically focused on machine learning fairness. Seven papers consisted of experience reports shared by researchers and practitioners, five papers proposed solutions to the problem of fairness, three evaluated them, and one discussed the philosophical issues concerned with fairness.

\begin{itemize}[nosep,leftmargin=0pt,labelindent=0pt]

    \item[\faBook]\hspace{.2pt} P1.1 -- The first identified practice is \textbf{\textit{Empirical requirements elicitation}}, which involves adopting empirical strategies such as surveys, interviews, focus groups, cognitive walkthroughs, and controlled experiments to elicit and validate fairness-oriented requirements under specific fairness constraints of the application. 
    \faInfoCircle \textit{Application.} Interview legal and domain experts to determine what fairness constraints are needed (e.g., features to base training on) to design the COMPAS system.

    \item[\faBook]\hspace{.2pt} P1.2 --The practice of \textbf{\textit{Multi-objective optimization}} should be used to balance different metrics in problem formulation, particularly considering ethical constraints. 
    \faInfoCircle \textit{Application.} Satisfy multiple mathematical constraints in COMPAS problem formulation to simultaneously minimize dependencies between model outcomes and skin color and increase the accuracy of the results as much as possible.

    \item[\faBook]\hspace{.2pt} P1.3 -- The last practice in the first category involves applying \textbf{\textit{reverse engineering}} approaches to existing products to elicit requirements for a fair-oriented solution under development. 
    \faInfoCircle \textit{Application.} Reconstruct the design diagrams of the COMPAS system (applying reverse engineering) to redesign it with the aim of mitigating the causes of discrimination.
    
\end{itemize}

\subsection{C2 - Data Preparation}

The second category of the catalog consists of seven practices related to data preparation for improving fairness in AI systems. The practices were identified in 55 papers, 39 of them specifically focused on the concept of fairness, and 16 generally treating the topic of machine learning.
A number of 18 papers consisted of experience reports shared by the authors, 25 papers proposed solutions to deal with fairness, seven papers evaluated existing approaches, four papers discussed about fairness from a phylosophical standpoint, and a single paper argued about personal opinions on the topic.

\begin{itemize}[nosep,leftmargin=0pt,labelindent=0pt]
    \item[\faBook]\hspace{.2pt} P2.1 -- The first practice in this category is \textbf{\emph{Data balancing techniques}}, which involves adopting specific strategies and methods for data balancing, such as oversampling, undersampling, uniform or preferential sampling, data filtering, labeling, or similar approaches, to reduce the causes of discrimination in a dataset. 
    \faInfoCircle \textit{Application.} Use oversampling to increase the number of black people samples to balance the dataset with respect to the skin color feature.

    \item[\faBook]\hspace{.2pt} P2.2 -- The second practice related to \emph{Data Preparation} groups \textbf{\emph{Data mining approaches}} that can be used to find discrimination within datasets, e.g., discriminative historical decision, not predefined minority group, or to analyze the semantic meaning of sensitive features. 
    \faInfoCircle \textit{Application.} Use manual or automatic data mining approaches to find new data that balance the number of black and white people in the COMPAS model dataset.

    \item[\faBook]\hspace{.2pt} P2.3 -- The third practice identified in this category involves \textbf{\emph{Data \& features transformation}} to ensure compliance with specific fairness constraints. This can include techniques such as dimensionality reduction, probabilistic transformation, matrix factorization, or data imputation. 
    \faInfoCircle \textit{Application.} Apply data-imputation strategies to the COMPAS model to mitigate bias due to the unavailability of data in minority groups (e.g., inferring from the ethnicity of individuals).

    \item[\faBook]\hspace{.2pt} P2.4 -- The fourth practice in this category is \textbf{\emph{Diversity data set selection}}. 
    This practice involves editing the dataset based on the similarity and diversity of data and features, with the goal of improving the representativeness of sensitive groups and avoiding discrimination among them. Examples of how this practice can be implemented include clustering algorithms or ranking algorithms.
    \faInfoCircle \textit{Application.} Use a ranking or clustering algorithm to identify minority groups (more likely to be discriminated against) concerning skin color.

    \item[\faBook]\hspace{.2pt} P2.5 -- The fifth practice in the \emph{Data Preparation} category is \textbf{\emph{Causal analysis approaches}}, which involves using causal graph approaches to identify and mitigate the causes of discrimination in datasets. These approaches aim to balance features, utility, and fairness of the data. 
    \faInfoCircle \textit{Application.} Use manual or automatic causal graph generation approaches for the COMPAS model to assess any dependence between model results and sensitive features.

    \item[\faBook]\hspace{.2pt} P2.6 -- The sixth practice in the catalog focuses on \textbf{\emph{Data fairness measurement}} and involves applying various measurement optimization strategies, such as data coverage, fairness metrics, and individual weights calculation. These strategies aim to monitor, analyze, or assess the fairness levels of data before training, considering different constraints like fairness and privacy. 
    \faInfoCircle \textit{Application.} Evaluate different privacy and fairness metrics on the COMPAS dataset to understand if the excessive presence of anonymous data can cause discrimination against people of a specific ethnicity.

    \item[\faBook]\hspace{.2pt} P2.7 -- The last practice in the second category of the catalog is \textbf{\emph{Multitask learning improvement}}, which involves learning strategies aimed at maximizing the average accuracy for each minority group in the training sampling. 
    \faInfoCircle \textit{Application.} Apply multitasking learning strategies to simultaneously improve levels of fairness and accuracy of outcomes toward black people in the context of the COMPAS model.
\end{itemize}

\subsection{C3 - Model Building}
The third category of the catalog groups four practices that are meant to be applied during the design and building phases of the model. Practices in this category have been derived from 44 papers, of which 27 proposed solutions to address fairness issues, two evaluated existing proposals, three discussed the concept of fairness from an ethical and philosophical standpoint, and 12 papers reported the authors' experiences in dealing with fairness.

\begin{itemize}[nosep,leftmargin=0pt,labelindent=0pt]

    \item[\faBook]\hspace{.2pt} P3.1 -- The first practice in the third category is named \textbf{\emph{Ensemble Learning}}, and consists of the use of meta-algorithms or ensemble learning approaches to guarantee high fairness, according to different fairness definitions and constraints. 
    \faInfoCircle \textit{Application.} Combine the outcomes of two learning strategies: one aimed at ensuring appropriate prediction outcomes for black people, the other at reducing the dependence between results and skin color.

    \item[\faBook]\hspace{.2pt} P3.2 -- The second practice in this category has been the most referenced of all in the collected papers; it was mentioned in 27 research works. It consists of \textbf{\emph{Focused Learning Strategies}}, such as clustering-based learning algorithms, that satisfy specific fairness definitions and constraints to achieve discrimination-free outcomes by balancing dependencies between sensitive and non-sensitive features. 
    \faInfoCircle \textit{Application.} Use learning approaches based on clustering analysis in order to produce targeted output results for each minority group identifiable by skin color.

   \item[\faBook]\hspace{.2pt} P3.3 -- The third practice in the third category consists of \textbf{\emph{Fair Regularization}}, which involves strategies to calibrate and optimize the model's loss function and other learning parameters by applying specific regularization factors calculated based on fairness metrics or following the distribution of training data.
   \faInfoCircle \textit{Application.} Modify the training function of the COMPAS model to mitigate and decrease the dependence on skin color with predictions of recidivism (adding a regularization term).

    \item[\faBook]\hspace{.2pt} P3.4 -- The last practice in the context of model building involves adopting \textbf{\emph{Adversarial Learning}} strategies to maximize prediction accuracy and minimize the probability of producing unfair outcomes from the model. 
    \faInfoCircle \textit{Application.} Apply adversarial learning strategies to balance fairness levels and outcome accuracy toward black people in the context of the COMPAS model.

\end{itemize}

\subsection{C4 - Model Training \& Testing}

The fourth category in the catalog focuses on strategies to be applied during model training and testing. A total of 38 papers were included in this category, with 28 of them specifically addressing fairness. Most of these papers aimed to propose solutions for addressing fairness issues, resulting in 25 solution proposals. Additionally, ten papers provided experience reports from researchers and practitioners.

\begin{itemize}[nosep,leftmargin=0pt,labelindent=0pt]

    \item[\faBook]\hspace{.2pt} P4.1 -- The first practice in this category involves applying \textbf{\emph{hyper-parameter tuning}} strategies to optimize the fairness levels of prediction results, such as using proper global or local search strategies or adding fair-regularization terms. 
    \faInfoCircle \textit{Application.} Apply global search hyper-parameters tuning to mitigate positive results among black and white people.

    \item[\faBook]\hspace{.2pt} P4.2 -- The second practice to be applied during the phase of model training and testing is the \textbf{\emph{Post-processing transformation}}, aimed at obtaining favorable outcomes for unprivileged groups using specific post-processing algorithms or strategies. These transformations include binary label transformations, score balancing between different sensitive groups, or community detection methods. 
    \faInfoCircle \textit{Application.} Mitigate COMPAS model outcomes (using transformation techniques) among groups of individuals with different skin colors but similar crime data.

    \item[\faBook]\hspace{.2pt} P4.3 -- The third practice consists of applying strategies of \textbf{\emph{post-processing optimization}}, such as matrix factorization or recommendation strategies, to evaluate and optimize learning outcomes based on given fairness metrics and constraints. 
    \faInfoCircle \textit{Application.} Apply weighted matrix factorization in order to mitigate COMPAS model results among people of different ethnicities.

    \item[\faBook]\hspace{.2pt} P4.4 -- The fourth practice in the fourth category is called \textbf{\emph{Test suite generation}} and involves the creation of specific fair test suites using various AI-based search or analysis techniques. These techniques include random sampling, global or local search, and symbolic execution, which aim to identify the discriminatory nature of a learner under test based on specific fairness definitions and/or potential minority groups. 
    \faInfoCircle \textit{Application.} Use automated sampling or search approaches to generate test suites to identify for which ethnicities of individuals the COMPAS model produces more unfavorable predictions.

    \item[\faBook]\hspace{.2pt} P4.5 -- The fifth practice in this category involves applying \textbf{\emph{Mutation testing}} strategies to identify possible causes of unfairness. These strategies include fine-grained data transformation techniques, such as changing features' names or editing features' values. 
    \faInfoCircle \textit{Application.} Generate mutant test samples to verify the presence of discrimination between individuals with different skin colors and similar non-sensitive features.

    \item[\faBook]\hspace{.2pt} P4.6 -- The last practice in the fourth category involves \textbf{\emph{Testing prediction oracles}} while considering fairness constraints. This is done by using positive/negative prediction tags as expected results for specific test inputs. 
    \faInfoCircle \textit{Application.} Adopt oracle testing strategies to verify that the prediction results of criminal recidivism match the expected results regardless of skin color.

\end{itemize}

\subsection{C5 - Model Verification \& Validation}

The fifth category of the catalog includes five practices that are meant to be applied during the phases of model verification and validation. These practices were mentioned in a total of 51 papers. Out of these, 28 discussed possible solutions, 10 shared experience reports, eight consisted of evaluation research, four argued philosophical aspects of software fairness, and one presented opinions on the concept.

\begin{itemize}[nosep,leftmargin=0pt,labelindent=0pt]

    \item[\faBook]\hspace{.2pt} P5.1 -- The first practice in this category is called \textbf{\emph{Meaning validation strategies}} and consists of reasoning on the sociological or linguistic meaning of data (or other information retrieval-related aspects) to detect the presence of discriminatory bias, for example, by using specific NLP or explainable-AI algorithms. 
    \faInfoCircle \textit{Application.} Create an approximate COMPAS model (with explainable AI strategies) to identify which training features might be influencing prediction to produce unfair outcomes.

    \item[\faBook]\hspace{.2pt} P5.2 -- The second practice related to model verification and validation consists of \textbf{\emph{Causal dependencies analysis}}. This practice aims to identify and eliminate potential causes of discrimination by examining the relationships between features. Examples include fair anomaly detection using directed acyclic graphs, clustering algorithms, and fairness-oriented data-slicing. 
    \faInfoCircle \textit{Application.} Apply K-Means for anomaly detection in the presence of discrimination between sensitive groups arising from skin color.

    \item[\faBook]\hspace{.2pt} P5.3 -- The third practice in the fifth category is \textbf{\emph{Model comparisons}}, which allows for the evaluation of fairness levels in a model by comparing it with other AI-intensive solutions. This can be done by using standard algorithms or alternative learning approaches that leverage similar parameters and learning configurations. 
    \faInfoCircle \textit{Application.} Compare the COMPAS model outcomes with another recidivism prediction model trained on the same dataset.

    \item[\faBook]\hspace{.2pt} P5.4 -- The fourth practice in this category consists of applying \textbf{\emph{Specific validation strategies}} that have to be defined ad-hoc, according to multiple fairness definitions and metrics, e.g., fairness probability measures or group similarity, or re-balancing strategies. 
    \faInfoCircle \textit{Application.} Identify (according to specific metrics) a minimum level of fairness that the COMPAS module must respect for each group of individuals identified by skin color.

    \item[\faBook]\hspace{.2pt} P5.5 -- The last practice in the fifth category groups \textbf{\emph{Formal validation strategies}}, e.g., leave-one-out cross-validation or regression strategies for testing, meant to evaluate fairness levels of the system and evaluate qualitative trade-offs.
    \faInfoCircle \textit{Application.} Apply Leave-One-Out cross-validation to test how accurate the model is in calculating the probability of committing theft among individuals of different ethnicities.

\end{itemize}

\subsection{C6 - Model Maintenance \& Evolution}

The final category of practices in the catalog pertains to the continuous phases of model maintenance and evolution, encompassing three practices. Six papers addressed practices within this category, with five of them specifically focusing on fairness. Among these papers, four proposed solutions, while two shared experience reports from practitioners and researchers.

\begin{itemize}[nosep,leftmargin=0pt,labelindent=0pt]

    \item[\faBook]\hspace{.2pt} P6.1 -- The first practice in the last category is \textbf{\emph{Feature standardization}}, which involves applying techniques such as linear or non-linear transformations to enhance the fairness levels of the AI-intensive solution. 

    \faInfoCircle \textit{Application.} Transform the numeric values of the sensitive feature in the COMPAS dataset to mitigate the risk of biases.

    \item[\faBook]\hspace{.2pt} P6.2 -- The second practice in the sixth category is \textbf{\emph{Model outcomes analysis}}, which involves evaluating and determining how to improve the fairness levels of the system after deployment. 

    \faInfoCircle \textit{Application.} Monitor and collect the actual prediction results of the COMPAS model, editing the unfair results for black people in order to retrain the model on the modified data.

    \item[\faBook]\hspace{.2pt} P6.3 -- The last practice in our catalog is called \textbf{\emph{Multiple datasets analysis}}, which involves enhancing the representativeness and fairness of the system by gathering data from multiple datasets. 

    \faInfoCircle \textit{Application.} Redesign the COMPAS training dataset by agglomerating multiple data sources to balance the number of individuals in different ethnic groups.

\end{itemize}

\definitionbox{{\faStickyNote} Summary of our Results}{Data Preparation emerged as the most significant category, with 7 practices mentioned in 55 papers, 39 of which specifically focused on fairness. Most categories showed a similar pattern: a majority of papers were solution-oriented, aiming to address fairness issues rather than just identifying them. Each category revealed unique approaches to mitigating bias: from empirical requirements gathering and data balancing in early stages, to ensemble learning and adversarial learning during model building, to meaning validation and causal dependency analysis during verification. Notably, our results suggest that fairness is not a one-time fix but requires ongoing attention. The practices demonstrate that addressing bias is a multi-dimensional challenge requiring technical, analytical, and contextual strategies at every stage of AI system development.}
\section{Discussions and Implication}
\begin{figure*}
        \centering
    \includegraphics[width=.65\linewidth]{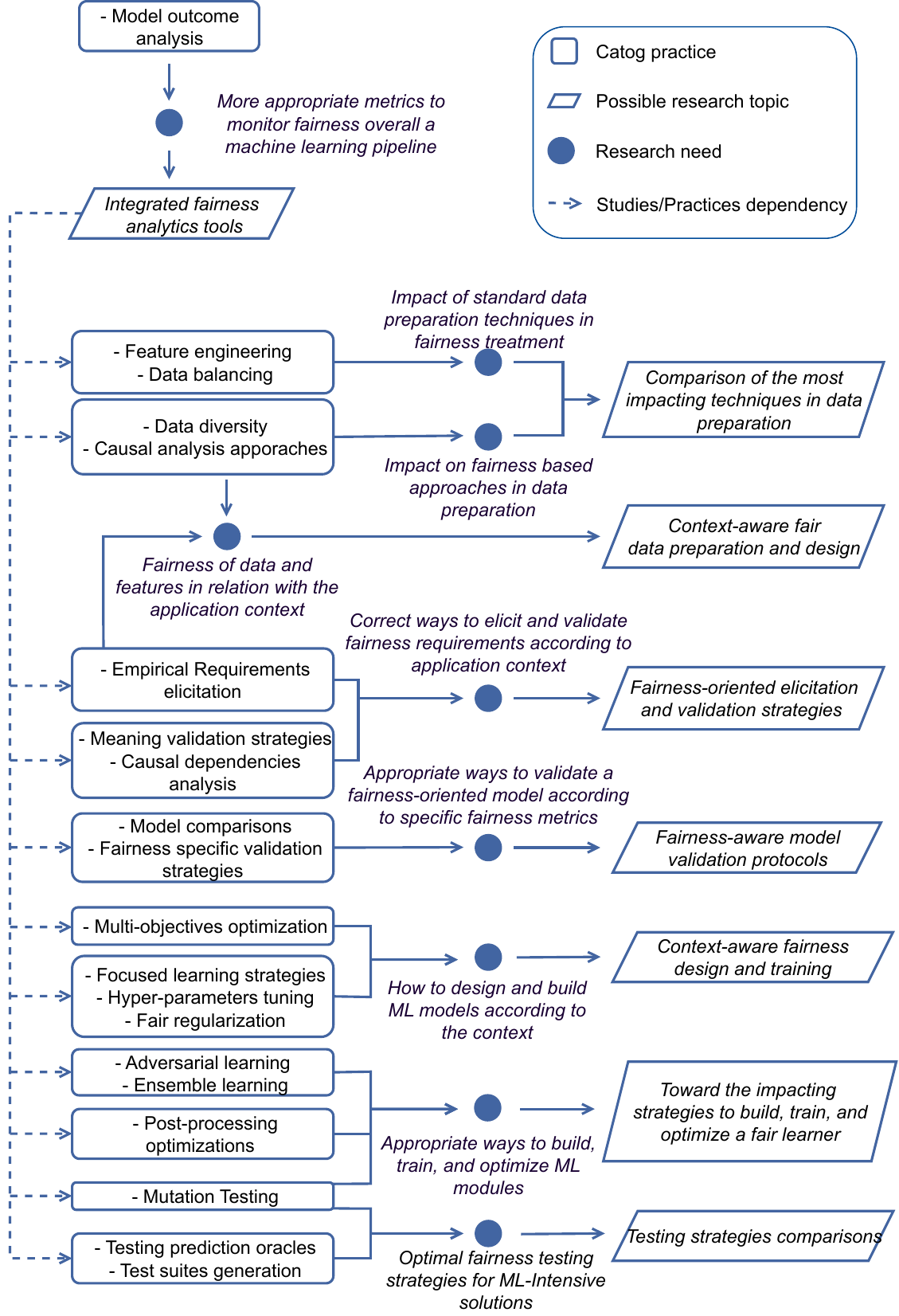}
        \caption{Research roadmap on Fairness-aware development and recommendations.}
        \label{im_a_roadmap}
\end{figure*}

In this section, we discuss the implications of our work. Our considerations are accompanied by a \textit{research roadmap} presented in Figure \ref{im_a_roadmap}, which summarizes the challenges emerging from our catalog. 

\smallskip \textbf{The relevance of fairness analytics and recommendations.}
Integrating fairness measurement tools throughout the AI development pipeline is crucial for ensuring the quality and reliability of AI systems. This systematic approach enables comprehensive evaluations across requirement analysis, data assessment, model optimization, and validation strategies. Based on our findings, researchers can now design recommendation systems that suggest and support the development of fairer AI solutions by leveraging the identified practices across different stages of AI development. Developing robust fairness monitoring tools can support researchers and practitioners in addressing potential biases, ultimately enhancing user trust and system transparency.

\smallskip \textbf{The tight dependency of fairness on training data.}
Our research underscores that fairness is fundamentally shaped by training data quality, with data preparation emerging as the most critical phase for implementing fairness-related practices. Traditional techniques like data balancing and feature engineering are essential in mitigating algorithmic biases, while emerging approaches such as data diversity selection and causal analysis offer promising new research directions for preventing systematic discrimination.

\smallskip \textbf{Context-aware fairness development.}
Ethical AI considerations must be tailored to specific application domains, necessitating context-aware fairness strategies. By integrating multi-objective optimization, fairness-aware data selection, and focused learning approaches, developers can create more equitable AI systems. Domain-specific validation strategies further ensure that AI-driven decision-making aligns with societal expectations and ethical standards.

\smallskip
\revised{\textbf{Fairness-aware AI assistants.} The insights from our research roadmap can be instrumental in enhancing fairness in AI-based assistants, which are increasingly used in decision-making, customer support, and personalized recommendations. Ensuring fairness in these systems requires integrating fairness-aware analytics throughout their lifecycle—from requirement elicitation to post-deployment monitoring. For instance, AI assistants can benefit from fairness-driven data preparation techniques, such as data diversity selection and causal analysis, to reduce biases in training datasets. Additionally, real-time fairness assessment tools could be embedded into AI assistants to continuously evaluate and mitigate bias in their responses or recommendations. By incorporating adaptive fairness-aware recommendation mechanisms, AI assistants can dynamically adjust their outputs based on evolving fairness constraints, ensuring equitable interactions across different user groups. Future research should explore automated mechanisms that allow AI-based assistants to detect and correct fairness issues autonomously, reinforcing trust and transparency in their deployment.}
\section{Limitations and Threats to Validity}
Multiple aspects might have influenced the research methods applied and the conclusions drawn to extract the catalog of fairness-aware practices.  
As for the systematic search process, the major challenge was related to the completeness of the keywords used within the search string. In this respect, we approached the problem by means of (1) some preliminary considerations on the current state of the art and (2) the inclusion of specific keywords along with more general terms typically used in the context of articles treating AI fairness. Perhaps more importantly, to be sure that our search process was representative of each phase of the AI development life cycle \cite{burkov2020machine}, we combined, using an \textsl{OR} operator, specific key words (representative of a specific MLOps phase), that helped us to obtain a reasonable initial set of papers to analyze.

In terms of filtering, we strictly adhered to the guidelines by Peterson \cite{petersen2015guidelines} and defined a set of exclusion and inclusion criteria to filter out non-relevant papers. In addition, we included a backward snowballing process with the aim of augmenting the set of relevant primary studies. Finally, before starting the data extraction and analysis, we conducted a formal resource classification step to verify the presence of the needed information in the resources.

Other possible threats to validity are concerned with the data extraction and analysis strategies that we adopted to formalize the catalog of fairness-aware practices. Following well-established guidelines \cite{petersen2015guidelines}, we defined a structured data extraction form and collected raw information that were later further analyzed to address our research goal.  Afterwards, to combine and summarize the results coming from the set of primary studies identified, we adopted a formal iterative content analysis \cite{white2006content} of the practices mentioned in those studies. This analysis included multiple iterations of micro-analysis and categorization, which were conducted by two inspectors to reduce risks due to subjective evaluations. In addition, the other authors where involved whenever needed to verify the consistency and validity of the methodical steps conducted. This formal procedure increases our confidence with respect to the soundness of the catalog elicited. Nonetheless, replications of our study conducted using different research methods (e.g., longitudinal investigations or industrial case studies) might reveal additional practices that we did not identify. To ease the work of other researchers and allow them to build on top of our catalog, we made all data available in our online appendix \cite{appendix}.

\section{Conclusions}
In this paper, we aimed at eliciting a catalog of fairness-aware practices that may be used to improve the levels of fairness within AI-enabled systems throughout the whole development life cycle. 
We conducted a systematic mapping study, followed by iterative content analysis sessions, to identify and validate a set of practices proposed in the literature. 
Our study reported 28 fairness-aware practices that span across six stages of AI development. 

The implications of the study propose an extensive roadmap for future research in the field of Software Engineering for Artificial Intelligence, Software Quality, and Empirical Software Engineering, and represent the input for our future research agenda. We first plan to further elaborate on the identified practices with empirical and software repository mining studies, focusing on measuring the impact of each practice on fairness metrics. Moreover, we intend to study novel methods and techniques to automate the application of fairness-aware practices, other than experimenting with strategies to support practitioners in improving the fairness of AI solutions.

\begin{acks}
We acknowledge the use of ChatGPT-4 to ensure linguistic accuracy and enhance the readability of this article. Further acknowledgments have been anonymized. This work has been partially supported by the European Union - NextGenerationEU through the Italian Ministry of University and Research, Project PRIN 2022 PNRR ``FRINGE: context-aware FaiRness engineerING in complex software systEms'' (grant n. P2022553SL, CUP: D53D23017340001). This work has been partially supported by the EMELIOT national research project, which has been funded by the MUR under the PRIN 2020 program (Contract 2020W3A5FY). This work was partially supported by project FAIR (PE0000013) under the NRRP MUR program funded by the EU - NGEU.
\end{acks}

\bibliographystyle{IEEEtran}
\bibliography{references}

\end{document}